\documentclass[aps,prl,twocolumn,amsmath,amssymb,reprint,superscriptaddress]{revtex4-1}

\usepackage[dvips]{graphicx}
\usepackage{multirow}
\usepackage{array}
\usepackage{bm}
\usepackage{amsmath}
\usepackage{threeparttable}
\usepackage{epstopdf}
\usepackage{threeparttable}
\usepackage{dcolumn}
\usepackage{ulem}
\usepackage[usenames,dvipsnames]{color}

\begin{document}

\title{The fracture strength of AlLiB$_{14}$}

\author{L. F. Wan }
\author{S. P. Beckman }
\email{sbeckman@iastate.edu.}

\affiliation{Department of Material Science and Engineering, Iowa State University, Ames, Iowa 50014, USA}%

\date{\today}

\begin{abstract}

The orthorhombic boride crystal family XYB$_{14}$, where X and Y are 
metal atoms, plays a critical role in a unique class of superhard 
compounds, 
yet there 
have been no studies aimed at understanding the origin of the 
mechanical strength of this compound. 
We present here the results from a comprehensive 
investigation into the fracture strength of the archetypal 
AlLiB$_{14}$ crystal. 
First-principles, \textit{ab initio}, methods are used to determine 
the ideal brittle cleavage strength for several high-symmetry 
orientations.
The elastic tensor and the orientation-dependent 
Young's modulus are calculated. 
From these results the lower bound fracture strength of 
AlLiB$_{14}$ is predicted to be between 29 and 31 GPa, which is 
near the measured hardness reported in the literature. 
These results indicate that the intrinsic strength of 
AlLiB$_{14}$ is limited by the interatomic B--B bonds that 
span between the B layers.

\end{abstract}


\maketitle

The development of new superhard materials that can operate under 
extreme conditions is critical for high-performance industrial manufacturing 
and is a subject that has recently received great attention.~\cite{Kaner,Solozhenko,McMillan} 
The orthorhombic borides, formulated as XYB$_{14}$ where X and Y are metal atoms, 
have been of interest to scientists and engineers for the past decade due to a 
report \cite{Cook2000,Lewis} that
AlMgB$_{14}$ prepared by mechanical milling can achieve a hardness 
between 32 and 46 GPa. The reason for the observed 
superhardness is not understood.  
It is suggested that in part the strength is due to the
so-called ``nano-composite'' microstructure comprised of 
AlMgB$_{14}$ and TiB$_{2}$, although the hardness of each 
individual phase is believed to be less than the hardness of the 
composite.~\cite{Cook2006} 
There have been many studies examining 
TiB$_{2}$,~\cite{Munro,Spoor,Okamoto,Heid} 
but the orthorhombic boride family has received much less attention 
and is therefore the focus of this letter. 

\begin{figure}[tbp]
\centering
\includegraphics[width=3in]{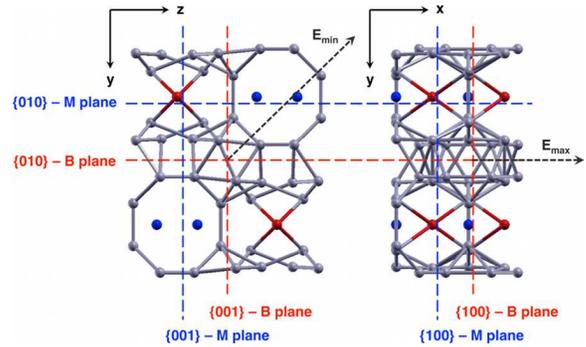}
\caption{(color online) A simple schematic of the XYB$_{14}$ crystal structure.  
The red (medium grey) spheres are the X-site, the blue (dark grey) spheres are the Y-site, 
and the light grey spheres are the B atoms.  The short-dashed arrows denoted 
by E$_{\mbox{min}}$ and E$_{\mbox{max}}$ show the loading directions that yield 
the minimum and maximum Young's modulus for AlLiB$_{14}$. The six planes 
selected for examination within the brittle cleavage model are shown as long-dashed 
lines and are labeled according to the nomenclature used in 
Table~\ref{table:cleaveresult}.}
\label{fig:cleaveplane}
\end{figure}

Whereas most hard materials are dense, highly-symmetric crystals, 
the XYB$_{14}$ structure, shown in Fig.~\ref{fig:cleaveplane}, 
is relatively open and has low symmetry (Imma). 
This crystal structure, which was first reported by Matkovich and Economy 
in 1970, has a unit cell containing four formula units of XYB$_{14}$.~\cite{Matkovich} 
The 64-atom unit cell can be expressed as 
X$_4$Y$_4$(B$_{12}$)$_4$B$_8$ to distinguish the two B allotropes.   
The B layers are constructed from B$_{12}$ icosahedron that 
are connected to each other through the so-called inter-icosahedra 
B atoms that are trigonally bonded to three B$_{12}$ units within the B layer. 
Recent spectroscopy evidence indicates that the B--B bonds that span 
between the B layers, directly connecting icosahedron, are very strong.~\cite{Werheit2010}
Unlike many metal-boride compounds the metal atoms are not covalently 
bonded to the B, but instead the metal atoms ionize and donate their valence 
electrons to the covalently bonded B network.~\cite{Harmon2002,Kolpin2008,Wan2012,Wan2011} 
As a near-superhard material, this crystal family is unique, which has led us to 
investigate the bonding in the crystal as it affects the crystal's mechanical strength.  

Following Matkovich and Economy's work,\cite{Matkovich} Higashi and Ito synthesized several 
XYB$_{14}$ compounds and used diffraction methods to refine the crystallographic data.  
For some of the compounds, such as AlMgB$_{14}$, a relatively high concentration of 
vacancies, around 25\%, are found at the metal atom sites.~\cite{Higashi1983}
Diffraction results for other XYB$_{14}$ compounds, 
such as AlLiB$_{14}$, do not find such a large number of 
vacancies.~\cite{Higashi1993,Peters2006}
Werheit \textit{et al.}~have used Raman spectroscopy to 
compared the vibrational spectrum of various XYB$_{14}$ 
compounds and have found that the AlLiB$_{14}$ crystal 
has less internal distortions than many other XYB$_{14}$ structures, 
including AlMgB$_{14}$.~\cite{Werheit2010} 
From these results we conclude that experimental specimens of 
AlLiB$_{14}$ are likely to have fewer point defects than many other 
XYB$_{14}$ compounds and consequently AlLiB$_{14}$ is selected as 
the archetypical structure for study in this letter.  

Previous theoretical studies have focused on the effects of chemical substitution 
on the properties of the XYB$_{14}$ crystal 
family.~\cite{Kolpin2008,Sahara2009,Harmon2002,Wan2012} 
This is in part because the system is known to accept a large number of metal species including 
Li, Be, Na, Mg, Al, as well as a variety of rare-earth elements, such as Tb, Dy, Ho, Er, Yb, and 
Lu.~\cite{Kudou2002,Okada2005,Higashi1993,Peters2006,Cook2000,Korsukova1989,Korsukova1993}
In addition the superhard Ames Lab specimen was synthesized by a mechanical 
alloying method, which introduces a wide variety of impurity species to the crystal including 
Ti, Si, Fe, O, and C.~\cite{Lewis} 
In these theoretical studies the reported figure of merit for hardness is the bulk modulus 
because its computation is relatively simple.
However, the bulk modulus alone only gives information about the average bond 
strength under an applied 
volume dilation and does not give any information about the strength of individual bonds in the crystal. 
Understanding hardness requires knowledge about the local mechanisms for bond breaking 
as it relates to fracture in the crystal. 
In this letter the fracture strength of AlLiB$_{14}$ will be examined using an ideal 
brittle cleavage model. 
This approach allows for insight regarding the local bonding within the crystal and may lead 
to a strategy for improving the hardness.  

The first-principles, density functional theory method used in this study is 
implemented in the SIESTA software package.~\cite{KS,SIESTA} 
The Perdew-Burke-Ernzerhof generalized gradient approximation (GGA) 
is used for the exchange-correlation energy and norm-conserving 
pseudopotentials are used in place of the all-electron atomic potentials.~\cite{PBE,TM} 
The wavefunction is represented by a set of finite-range numerical atomic orbitals. 
Each atomic basis is extended to include double$-\zeta$ functions plus a shell 
polarization that is constructed using the split-valence scheme.~\cite{Basis} 
The cutoff radii used for each $\zeta$ function are presented 
in Ref.~\onlinecite{waninprep2012}. 
Real space meshing is performed to an energy cutoff of 175 Rydberg. 
The Kohn-Sham energies are sampled across the Brillouin zone using 
a $12\times12\times12$ Monkhorst-Pack grid.~\cite{MP} 
The atomic structural optimization follows the conjugate gradient minimization 
method and the thresholds for the residual forces on atoms and the supercell 
are 0.005~eV/\AA~and 0.0005~eV/\AA$^3$ respectively. The calculated lattice 
parameters for AlLiB$_{14}$ are 5.88, 10.39, and 8.15~\AA, which agree well 
with the reported experimental values 5.847, 10.354, and 8.143~\AA.~\cite{Higashi1993} 

The ideal brittle cleavage model used here separates the AlLiB$_{14}$ crystal 
into two semi-infinite, rigid atomic blocks that are pulled apart to introduce a 
pair of cleavage surfaces at a predefined atomic plane. 
This idealized approach simultaneously 
stretches and breaks all the bonds at the interface. 
Although the effect of crack tip initiation and 
propagation cannot be included using this method, it 
allows for the bond strengths localized in the crystal to be 
investigated.  
Internal atomic relaxations and lattice contractions perpendicular 
to the direction of elongation are forbidden. 
These constraints allow the strength of the bonds across the 
cleavage interface to be determined independent of possible 
near-surface atomic reconstructions, which would be present in an experimental specimen. 
The calculated energy of the cleaved crystal, relative to the energy of the perfect crystal, is called 
the decohesive energy, $E_{b}$, and is determined as a function of the interplaner spacing, $x$, 
across the specified cleavage planes. The decohesive energy is fit to the 
universal binding energy relation (UBER) developed by Rose \textit{et al.}~in 
Ref.~\onlinecite{Rose}, which is expressed more precisely in Ref.~\onlinecite{Lazar2005} as, 
\begin{equation*}
E_b(x) = G_b \left[1-\left(1+\frac{x}{l_b}\right)exp\left(-\frac{x}{l_b}\right)\right] . 
\end{equation*}
When all the atomic bonds that span the cleavage interface are broken, 
the decohesive energy saturates to the cleavage energy, $G_b$. 
The cleavage stress is the first derivative of the decohesive energy with respect to $x$. 
The critical cleavage stress, $\sigma_{b}$, is defined as the maximum stress, 
and the corresponding interplanar spacing is referred to as the critical length, $l_{b}$.

\begin{figure}[btp]
\centering
\includegraphics[width=3in]{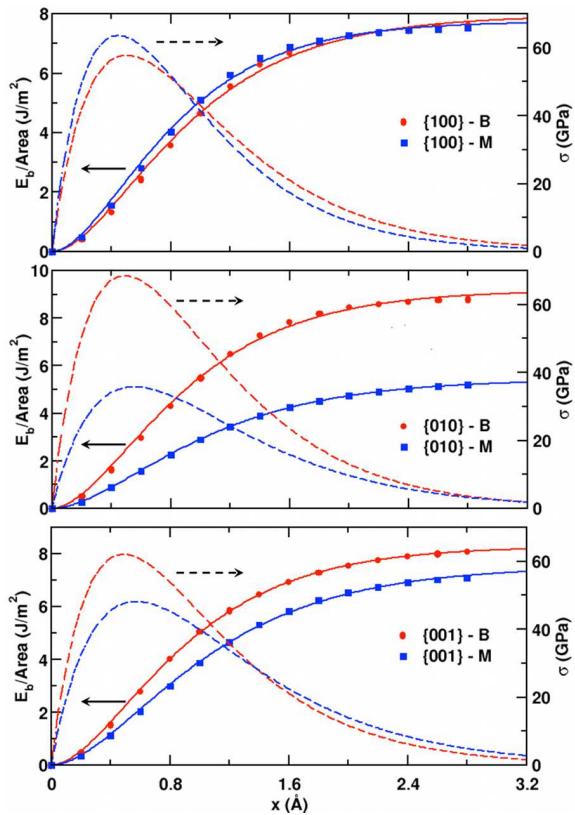}
\caption{(color online) The ideal brittle cleavage results for AlLiB$_{14}$. 
The left ordinate axis labels the energies and the right the stresses. 
The top frame gives the results for the \{100\} planes, 
the middle the \{010\} planes, and the bottom the \{001\} planes. 
In each frame the ``B'' results are red (light grey) and the ``M'' results 
are blue (dark grey). The decohesive energy DFT data are solid symbols, 
the UBER relations are solid lines, and the stresses are dashed lines.}
\label{fig:cleaveresult}
\end{figure}

For AlLiB$_{14}$ cleavage is considered within the high-symmetry \{100\}, \{010\}, and \{001\} 
families of planes. 
For each crystallographic direction two cleavage planes are examined: 
one that passes through the icosahedron, labeled ``B,'' and one that passes between the 
icosahedron, labeled ``M.'' 
These were selected to best represent the variation in the bonding for 
each of the sampled directions; one of the cleavage planes has many bonds 
that span the interface and the other few bonds. It is intended to 
test interfaces with the highest and lowest fracture energies.
These planes are identified in Fig.~\ref{fig:cleaveplane}. 
A 128-atom supercell is used to 
guarantee that the calculated decohesive energies are 
converged to better than 0.005 J/m$^2$. 
The decohesive energies, UBER fit, and derived stresses 
are plotted in Fig.~\ref{fig:cleaveresult}.  The computed data matches the functional 
form of the UBER relation very well and the resulting 
critical parameters are listed 
in Table~\ref{table:cleaveresult}. 

\begin{table}[tbp]
\begin{center}
\caption{\label{table:cleaveresult} The numerical results for applying the ideal brittle cleavage 
model to the six AlLiB$_{14}$ cleavage planes that are shown in Fig.~\ref{fig:cleaveplane}.}
\setlength{\tabcolsep}{2pt}
{\setlength{\extrarowheight}{6pt}
\begin{tabular}{c | c c c }
\hline 
\hline 
\multirow{2}{*}{Orientation} & Cleavage energy & Critical length & Critical stress \\
& $G_b$/Area (J/m$^2$) & $l_b$ (\AA) & $\sigma_b$ (GPa) \\ [1ex]
\hline
\{100\}--B & 7.94 & 0.51 & 57.7 \\
\{100\}--M & 7.74 & 0.45 & 63.5  \\ [1ex]
\hline
\{010\}--B & 9.16 & 0.49 & 68.4 \\
\{010\}--M & 5.42 & 0.56 & 35.7 \\ [1ex]
\hline
\{001\}--B & 8.27 & 0.49 & 62.0 \\
\{001\}--M & 7.51 & 0.57 & 48.1 \\ [1ex]
\hline 
\end{tabular}}
\end{center}
\end{table}

The decohesive energy curves and stresses 
for the \{100\}--B and --M planes are very similar, the critical stresses 
differ by less than 10\%.  This is not surprising considering that the density 
and geometric arrangements of the B--B bonds in these planes are nearly 
equivalent.  
Whereas the \{100\} planes are very similar, the 
\{010\} planes are considerably different.  
The \{010\}--B plane passes through the B layer bisecting the icosahedron, 
breaking many B--B bonds, but the \{010\}--M plane passes between the 
B layers and therefore cuts significantly fewer bonds. 
Within the ideal brittle cleavage model the calculated critical stress for the M plane is 
48\% smaller than that of the B plane. 

\begin{figure}[btp]
\centering
\includegraphics[width=3in]{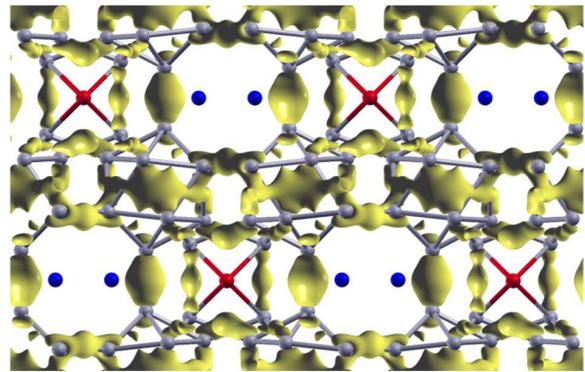}
\caption{(online color)
An isosurface showing regions in an XYB$_{14}$ crystal with positive bonding charge density. 
The atomic sites are color coded following the description in Fig.~\ref{fig:cleaveplane}.
} 
\label{chargeden}
\end{figure}

According to the Raman spectroscopy results reported in 
Ref.~\onlinecite{Werheit2010}, the B--B bonds that span between the B layers and connect 
the icosahedron are 
expected to have a greater binding strength than the bonds inside the icosahedron. 
This can be examined qualitatively by plotting 
the bonding charge density, as shown in 
Fig.~\ref{chargeden}.~\footnote{The bonding charge density is rendered by 
subtracting the charge density of the bonded structure from the sum of 
the charge contribution from each of the atoms, treated as though they 
are isolated. A positive bonding charge density indicates charge accumulation 
and a negative value indicates charge depletion.} 
The bonding process results in a buildup of charge in the   
B--B bonds that bridge the B layers. 
In Fig.~\ref{chargeden} charge accumulation between the B layers 
is observed both between the icosahedron and the inter-icosahedra B.
The results from the ideal cleavage model, presented in 
Table \ref{table:cleaveresult}, indicate that regardless of the anticipated high 
strength of the bonds at this plane, the relatively low number density of bonds  
causes the \{010\}--M plane to have the lowest critical stress 
of all the planes examined here.
It can be concluded that the B--B bonds that span between the B layers 
are key for controlling the overall strength of the crystal. 

\begin{figure}[tbp]
\centering
\includegraphics[width=3.5in]{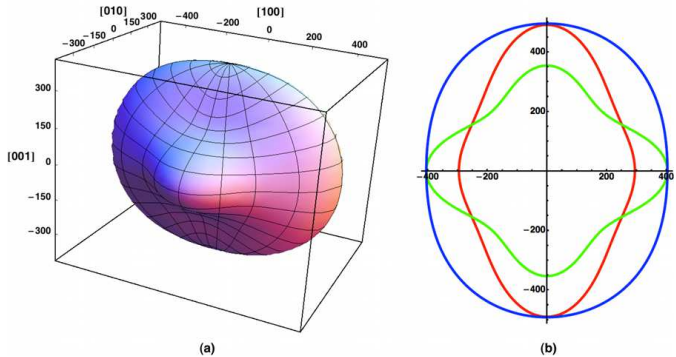}
\caption{(color online) The representation surface for the Young's modulus of AlLiB$_{14}$ (given in GPa) as a function of 
crystallographic orientation. Frame (a) shows a 3D plot of the representation surface. 
Frame (b) shows 2D projections of the representation surface onto the 
(100), (010) and (001) planes and are printed in green (light grey), 
blue (dark grey), and red (medium grey) respectively.}
\label{fig:elastic}
\end{figure}

It is the weakest planes that are of primary 
interest because fracture naturally transverses  
the weakest path through a crystal. 
It is these same planes that also dominate the elastic 
response.
For a given family of planes, the elastic response to a uniaxial load 
applied normal to the planes should be an indicator of the relative 
cleavage strength, \textit{i.e.}, the Young's modulus should scale with the 
cleavage strength. 
For the orthorhombic XYB$_{14}$-type crystal there are nine 
unique tensor elements that can be derived from the linear stress-strain relation 
and the crystal symmetries. 
The components of the stiffness and compliance tensors for AlLiB$_{14}$ are 
calculated and presented in Table~\ref{table:elastic}. 
In Fig.~\ref{fig:elastic}(a), the Young's modulus, $E$, is represented as a 
function of crystallographic orientation, according to the formula,
\begin{align*}
\frac{1}{E}  &=  l_1^{\hfill{4}}s_{11} + l_2^{\hfill{4}}s_{22} + l_3^{\hfill{4}}s_{33} + 2l_1^{\hfill{2}}l_2^{\hfill{2}}s_{12} + 2l_1^{\hfill{2}}l_3^{\hfill{2}}s_{13} + 2l_2^{\hfill{2}}l_3^{\hfill{2}}s_{23} \\ \nonumber
 &+  l_2^{\hfill{2}}l_3^{\hfill{2}}s_{44} + l_1^{\hfill{2}}l_3^{\hfill{2}}s_{55} + l_1^{\hfill{2}}l_2^{\hfill{2}}s_{66} ,\nonumber
\end{align*}
where $s_{ij}$ are the elastic compliance tensor components 
and $l_1$, $l_2$, and $l_3$ are the direction cosines. 
The representation surface in Fig.~\ref{fig:elastic}(a) is projected 
on the (100), (010) and (001) planes, and the results are shown 
in Fig.~\ref{fig:elastic}(b).

\begin{table*}[tbp]
\begin{center}
\caption{\label{table:elastic} The elastic tensor components for AlLiB$_{14}$.}
\setlength{\tabcolsep}{4pt}
{\setlength{\extrarowheight}{6pt}
\begin{tabular*}{120mm}{c | c c c c c c c c c}
\hline 
\hline 
Stiffness coefficients & c$_{11}$ & c$_{22}$ & c$_{33}$ & c$_{44}$ & c$_{55}$ & c$_{66}$ & c$_{12}$ & c$_{13}$  & c$_{23}$ \\
(GPa) & 526  & 411 & 419 & 91.0 & 201 & 130 & 45.7 & 83.4 & 32.0 \\ [1ex]
\hline
Compliance coefficients & s$_{11}$ & s$_{22}$ & s$_{33}$ & s$_{44}$ & s$_{55}$ & s$_{66}$ & s$_{12}$ & s$_{13}$  & s$_{23}$ \\
($\times$10$^{-12}$m$^2$/N) & 1.98  & 2.47 & 2.47 & 10.99 & 4.98 & 7.69 & -0.19 & -0.38 & -0.15 \\ [1ex]
\hline 
\end{tabular*}}
\end{center}
\end{table*}

For the [010], [001], and [100] directions the Young's modulus is  
293, 404 and 505 GPa. From the ideal brittle cleavage model 
the minimum critical stresses for these same directions are 
35.7, 48.1, and 57.7 GPa, as listed in Table \ref{table:cleaveresult}. 
Comparing these numbers demonstrates that indeed the 
directional representation of the Young's modulus is 
an accurate predictor of the relative ideal fracture strength of a 
particular orientation. 
From the results in Fig.~\ref{fig:elastic} the minimum value of 
the Young's modulus is 256.0 GPa, which corresponds to a uniaxial load 
orientated 
$\left(\phi= 90^{\circ} \mbox{, }\theta=44.96^{\circ}\right)$, where 
$\phi$ is the angle of rotation from the positive $x-$axis to the positive 
$y-$axis in the $xy-$plane and $\theta$ is the out-of-plane angle of rotation 
from the positive $z-$axis to the $xy-$plane.
This direction is drawn in Fig.~\ref{fig:cleaveplane} as a short-dashed arrow 
that is labeled E$_{\mbox{min}}$.
Assuming linear proportional scaling, the computed results for the 
high-symmetry orientations can be used to predict that the cleavage strength 
for a uniaxial load applied in the E$_{\mbox{min}}$ direction is between 
29 and 31 GPa. 
This is the predicted lower limit of the ideal brittle cleavage strength 
for AlLiB$_{14}$.  
We submit that for a brittle material, such as AlLiB$_{14}$, 
which does not undergo extensive plastic deformation near the crack tip prior to fracture, 
the calculated ideal brittle cleavage strength is a reasonable estimation of the fracture strength.
The physical features of fracture neglected in this ideal brittle cleavage model, 
including crack tip plasticity, lattice contractions, and atomic reconstructions, results in an overestimation 
of the cleavage energy and subsequently the actual critical energy and stress will be lower than our 
calculated results.
The experimentally measured hardness for AlLiB$_{14}$ ranges between 
20 and 29~GPa~\cite{Higashi1993,Kudou2002}, which suggests that for this 
material the atomic scale behavior that we have reported here plays an important 
role in determining the actual hardness of the material.  

In summary, we have coupled the results from a series of ideal brittle cleavage strength 
calculations to the calculated orientation dependent Young's modulus to predict the 
fracture strength of AlLiB$_{14}$.  
While admittedly this simplistic model neglects some of the macroscopic features 
of fracture associated with crack-tips, lattice plasticity, and interface reconstructions, 
we believe that here we have demonstrated that this still may be an effective approach to 
gauge the strength of brittle materials, such as the XYB$_{14}$ crystal family. 
In contrast to all 
of the previous theoretical studies of the XYB$_{14}$ crystal family, 
which have used the bulk modulus as an indicator of the bond strength, the approach used 
here allows for the local bond strength to be investigated on a plane-by-plane basis.  
Unlike the more sophisticated, multi-scale modeling approaches, which have been 
deployed to study fracture in polycrystalline diamond, Si, and other metallic systems,~\cite{Yang,Kaxiras,Beltz,Miller}
the method used here is relatively simple. 
We believe that our approach can be used to screen prospective structures prior to 
their being investigated using a more elaborate theoretical technique.  

The existing picture of bonding in the XYB$_{14}$ crystal family is that
B forms a covalent network of atoms constructed of B${12}$ icosahedron. The B$_{12}$ 
are stabilized by the electrons donated by the ionized metal atoms, according to the 
Jemmis mno rules.~\cite{Jemmis}
Excess charge accumulates in the inter-icosahedra bonds, both those within the B layer 
and those that span between the layers.  Experimental results indicate that the 
inter-icosahedra B--B bonds spanning between the layers are stronger than the 
intra-icosahedra bonds.~\cite{Werheit2010} 
Here we find that regardless the strength of the inter-icosahedra bonds the fracture is 
significantly more likely to proceed between the icosahedron rather than through, due to 
the density of bonds at the cleavage plane. 
In fact the \{010\}--M planes are the weakest of those examined in this study, 
which suggests that the hardness of the material may be closely tied to the B--B 
bonding that connects the icosahedra layers. 
In practice, this means the intrinsic strength of this crystal family possibly can be 
changed, either strengthened or weakened, by the introduction of a dopant 
species that directly affects these bonds. 

The authors gratefully acknowledge financial support from the National Science Foundation 
through grant DMR-1105641.


%

\end{document}